\documentclass[aps,prc,showpacs]{revtex4}

\usepackage{epsfig}

\newcommand{\vsig}{\mbox{\boldmath$\sigma$\unboldmath}}

\newcommand{\be}{\begin{equation}}
\newcommand{\ee}{\end{equation}}
\newcommand{\bea}{\begin{eqnarray}}
\newcommand{\eea}{\end{eqnarray}}
\newcommand{\bean}{\begin{eqnarray*}}
\newcommand{\eean}{\end{eqnarray*}}

\newcommand{\gapproxeq}{\lower
.7ex\hbox{$\;\stackrel{\textstyle >}{\sim}\;$}}
\newcommand{\lapproxeq}{\lower
.7ex\hbox{$\;\stackrel{\textstyle <}{\sim}\;$}}

\begin{document}

\title{Study of $\pi^{-}p \rightarrow \eta n$ at low energies in a chiral constituent quark model }
\author{
Xian-Hui Zhong$^1$ and Qiang Zhao$^{1,2}$}

\affiliation{ 1) Institute of High Energy Physics,
       Chinese Academy of Sciences, Beijing 100049, P.R. China
}
\affiliation{ 2) Department of Physics, University of Surrey,
Guildford, GU2 7XH, United Kingdom
            }

\author{  Jun He and Bijan Saghai }
\affiliation{ Laboratoire de recherche sur les lois fondamentales
de l'Univers, DAPNIA/SPhN, CEA/Saclay, 91191 Gif-sur-Yvette,
France }

\begin{abstract}

A chiral quark model approach is extended to the study of the $\pi
N$ scattering at low energies. The process of $\pi^{-}p
\rightarrow \eta n$ near threshold is investigated. The model is
successful in describing the differential cross sections and total
cross section near the $\eta$ production threshold. The roles of
the resonances in $n\leq 2$ shells are clarified. Near threshold,
the $S_{11}(1535)$ dominates the reactions, while the
interferences from the $S_{11}(1650)$ turn out to be destructive
around $W\lesssim 1.6$ GeV. The $D_{13}(1520)$ is crucial to give
correct shapes of the differential cross sections. The nucleon
pole term contributions are significant. The $P_{11}(1710)$ plays
an important role around the c.m. energy $W=1.7$ GeV, it is
crucial to produce an enhancement in the region of $W>1.6$ GeV as
suggested by the data for total cross section. The $t$-channel is
negligible in the reactions.
\end{abstract}



\maketitle

\section{introduction}

The $\pi^{-}p \rightarrow \eta n$ reaction at low energies is an
interesting topic in nuclear physics. This reaction can provide a
good probe into the structure of some low-lying resonances, such
as $S_{11}(1535)$, of which the property still bares a lot of
controversies. In the naive quark model, it is classified as the
lowest $L=1$ orbital excited state with $J^P=1/2^-$. Recently, it
is argued that it may contain a large admixture of pentaquark
component~\cite{Liu:2005pm}, which will explain the reversed mass
ordering between the $S_{11}(1535)$ and $P_{11}(1440)$. By
studying this reaction, one can extract the $\eta N$ interaction,
for which a possible strong attraction between $\eta$ and $N$ at
low energies may lead to ``$\eta$-mesic
nuclei"~\cite{eta-mesic,Baru:2006hy}. In general, more and more
accurate data from $\pi^{-}p \rightarrow \eta n$ experiments will
provide a challenging testing ground for the low energy theories
of hadron interactions, such as chiral perturbation theory,
meson-exchange model, etc.

On the process $\pi^{-}p \rightarrow \eta n$, there have been a
few experiments. The data come mainly from the old measurements
about thirty years ago \cite{exp1,exp2,exp3,exp4,exp5,exp6}, which
have been reviewed by Clajus and Nefkens \cite{Clajus:1992dh}.
Fortunately, a recent $\pi^{-}p \rightarrow \eta n$ experiment was
performed at BNL using the Crystal Ball spectrometer \cite{exp7}.
The differential cross sections together with total cross section
for $\eta$ production in reaction $\pi^{-}p \rightarrow \eta n$
have been measured at the incident $\pi$ beam momenta from
threshold to $p_\pi= 747 $ MeV/c. The quality of the data was
significantly improved compared with the previous measurements.
Theoretically, a few typical models have been used to deal with
the $\pi^{-}p \rightarrow \eta n$ reactions
\cite{Penner:2002ma,Shklyar:2004dy,th2,th3,th4,th5,th6,Vrana:1999nt},
such as the coupled-channel model, meson-exchange model, the
chiral multi-channel model. As pointed out in ref.~\cite{th1}, the
present theory is far from being as accurate as the experiment.
Thus, more theoretical studies are needed.

In this work, we introduce an effective chiral Lagrangian to
describe the quark-pseudoscalar-meson coupling and study the
meson-nucleon scattering in the constituent quark model. This
approach has been successfully applied to the study of the meson
photoproduction off nucleons
~\cite{qk1,qk2,qkk,qkk0,qkk1,qkka,Li:1997gda,qkk2,qk3,qk4,qk5,qk6}.
Since the quark-pseudoscalar-meson coupling is invariant under the
chiral transformation, some of the low-energy properties of QCD
are retained. There are several outstanding features for this
model. One is that only a very limited number of parameters will
appear in this framework. In particular, only one parameter is
need for the resonances to be coupled to the pseudoscalar mesons.
This distinguishes from hadronic models where each resonance
requires one additional coupling constant as free parameter. The
second is that all the resonances can be treated consistently in
the quark model. Thus, it has predictive powers when exposed to
experimental data, and information about the resonance structures
can be extracted.

However, it should be clarified that we restrict the quark-meson
interactions in the scattering processes where the mesons are
external fields interacting with the constituent quarks of the
Isgur-Karl model~\cite{isgur-karl-model}. Thus, the
spin-independent quark confinement potential is described by
harmonic oscillator potential. This allows an analytic separation
of the intermediate meson excitation matrix elements. In
principle, the quark-meson interaction will influence the
description of the constituent quark potentials, e.g.,
modifications to the quark interactions may occur and naive quark
model spectrum will be changed. We leave this to be investigated
in future development of this approach. The baryon spectroscopy
studied via Goldstone-boson exchanges (GBE) can be found in
Ref.~\cite{glozman}. Extended chiral quark model approach
combining both one-gluon-exchange (OGE) and GBE potentials has
also been investigated in the
literature~\cite{zhangzy-97,huang-05}.

In this work, we have investigated the $\pi^{-}p \rightarrow \eta
n$ reaction from the $\eta$ production threshold to the  c.m.
energy $W\simeq 1.7$ GeV. Our results are in good agreement with
the data. We find that $S_{11}(1535)$ dominates the reaction
around threshold. The resonances $D_{13}(1520)$ and $S_{11}(1650)$
also play very important roles in the process. The $D_{13}(1520)$
is crucial to give correct shapes of the differential cross
sections, although its contributions to the cross section are very
small near threshold. The $S_{11}(1650)$ has important destructive
interferences with the dominant $S_{11}(1535)$ around $W\lesssim
1.6$ GeV. Above the c.m. energy $W\simeq 1.6$ GeV, the
contributions of higher resonances from $n=2$ shell also appear.
The predictions of differential cross sections become worse with
the increasing c.m. energy $W$. The resonance $P_{11}(1710)$ plays
an important role, it is crucial to produce an enhancement in the
region of $W>1.6$ GeV as suggested by the data for total cross
section, and with which the theoretical predictions are obviously
improved if we change the sign of its amplitude. The nucleon pole
term contributions turn out to be necessary though a relatively
small $g_{\eta NN}$ coupling is favored. The $t$-channel is
negligible in the reactions.

The paper is organized as follows. In the subsequent section, the
framework is outlined. Then, the transition amplitudes in the
quark model are derived in Sec.\ \ref{am}. The resonance
contributions are separated out in Sec.\ \ref{rs}. We present our
calculations and discussions in Sec.\ \ref{RD}. Finally, a summary
is given in Sec.\ \ref{sum}.

\section{framework}

In the chiral quark model, the low energy quark-meson interactions
are described by the effective Lagrangian \cite{Li:1997gda,qk3}
\begin{eqnarray} \label{lg}
\mathcal{L}=\bar{\psi}[\gamma_{\mu}(i\partial^{\mu}+V^{\mu}+\gamma_5A^{\mu})-m]\psi
+\cdot\cdot\cdot,
\end{eqnarray}
where $V^{\mu}$ and $A^{\mu}$ correspond to vector and axial
currents, respectively. They are given by
\begin{eqnarray}
V^{\mu} &=&
 \frac{1}{2}(\xi\partial^{\mu}\xi^{\dag}+\xi^{\dag}\partial^{\mu}\xi),
\nonumber\\
 A^{\mu}
&=&
 \frac{1}{2i}(\xi\partial^{\mu}\xi^{\dag}-\xi^{\dag}\partial^{\mu}\xi),
\end{eqnarray}
with $ \xi=\exp{(i \phi_m/f_m)}$, where $f_m$ is the meson decay
constant. For the SU(3) case, the pseudoscalar-meson octet
$\phi_m$ can be expressed as
\begin{eqnarray}
\phi_m=\pmatrix{
 \frac{1}{\sqrt{2}}\pi^0+\frac{1}{\sqrt{6}}\eta & \pi^+ & K^+ \cr
 \pi^- & -\frac{1}{\sqrt{2}}\pi^0+\frac{1}{\sqrt{6}}\eta & K^0 \cr
 K^- & \bar{K}^0 & -\sqrt{\frac{2}{3}}\eta},
\end{eqnarray}
and the quark field $\psi$ is given by
\begin{eqnarray}
\psi=\pmatrix{\psi(u)\cr \psi(d) \cr \psi(s) }.
\end{eqnarray}

From the leading order of the Lagrangian [see Eq.(\ref{lg})], we
obtain the standard quark-meson pseudovector coupling at tree
level
\begin{eqnarray}\label{coup}
H_m=\sum_j
\frac{1}{f_m}\bar{\psi}_j\gamma^{j}_{\mu}\gamma^{j}_{5}\psi_j\partial^{\mu}\phi_m.
\end{eqnarray}
where $\psi_j$ represents the $j$-th quark field in the nucleon.

\begin{center}
\begin{figure}[ht]
\centering \epsfxsize=8.0 cm \epsfbox{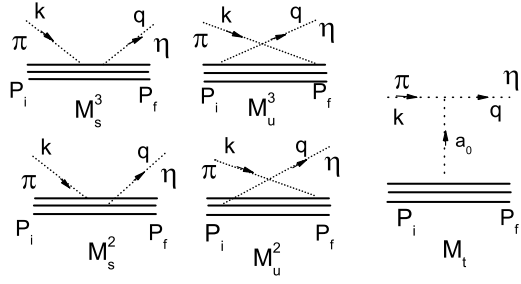}
\caption{\footnotesize $s$- $u$- and $t$-channels are considered
in this work. $M^{s}_{3}$ and $M^{u}_{3}$ ($M^{s}_{2}$,
$M^{u}_{2}$) correspond to the amplitudes of $s$- and $u$-channels
for the incoming meson and outgoing meson absorbed and emitted by
the same quark (different quarks), respectively.}\label{fig-1}
\end{figure}
\end{center}

The $\eta$ meson production amplitude (see Fig. \ref{fig-1}) can
be expressed in term of the Mandelstam variables:
\begin{eqnarray}
\mathcal{M}=\mathcal{M}_{s}+\mathcal{M}_{u}+\mathcal{M}_t \ .
\end{eqnarray}
The $s$- and $u$-channel transitions are given by
\begin{eqnarray}
\mathcal{M}_{s}=\sum_j\langle N_f |H_{\eta} |N_j\rangle\langle N_j
|\frac{1}{E_i+\omega_\pi-E_j}H_{\pi }|N_i\rangle, \label{sc}\\
\mathcal{M}_{u}=\sum_j\langle N_f |H_{\pi }
\frac{1}{E_i-\omega_\eta-E_j}|N_j\rangle\langle N_j
| H_{\eta} |N_i\rangle, \label{uc}
\end{eqnarray}
where $\omega_\pi$ and $\omega_\eta$ are the energies of the
incoming $\pi$-meson and outgoing $\eta$-meson, respectively.
$H_\pi$ and $H_\eta$ are the standard quark-meson couplings at
tree level described by Eq.(\ref{coup}). $|N_i\rangle$,
$|N_j\rangle$ and $|N_f\rangle$ stand for the initial,
intermediate and final states, respectively, and their
corresponding energies are $E_i$, $E_j$ and $E_f$, which are the
eigenvalues of the NRCQM Hamiltonian
$\hat{H}$~\cite{isgur-karl-model}.

 Following the procedures developed in refs.
\cite{qkk,Li:1997gda,qk3}, one can then express the $s$ and $u$
channel amplitudes by operator expansions. For instance, the $s$
channel can be written as \be \mathcal{M}_{s}=\sum_j\langle N_f
|H_{\eta} |N_j\rangle\langle N_j| \sum_n
\frac{1}{\omega_\pi^{n+1}}(\hat{H}-E_i)^n H_{\pi }|N_i\rangle \ ,
\ee where $n$ is the harmonic oscillator quantum number. Note that
for any operator ${\cal O}$, one has \be (\hat{H}-E_i){\cal
O}|N_i\rangle = [\hat{H}, \ {\cal O}]|N_i\rangle , \ee a
systematic expansion of the commutator between the NRCQM
Hamiltonian $\hat{H}$ and the vertex coupling $H_\pi$ and $H_\eta$
can thus be carried out. Details of this treatment can be found in
refs. \cite{qkk,Li:1997gda,qk3}, but we note that in this study
only the spin-independent potential in $\hat{H}$ is considered as
a feasible leading order calculation.

From PDG \cite{PDG} we know that the $a_0$ meson decay is
dominated by $\pi \eta$ channel. Thus, we consider $a_0$ exchange
as the dominant contributions to the $t$-channel transitions. For
the $\pi \eta a_0$ coupling, we introduce the following effective
Lagrangian
\begin{eqnarray}
\mathcal{L}_{a_0\pi \eta }&=&g_{a_0\pi \eta } m_\pi \eta \vec{\pi}
\vec{a}_0,
\end{eqnarray}
and for the quark-$a_0$ coupling, we select a scalar interaction
\begin{eqnarray}
H_{a_0}&=&\sum_j g_{a_0 qq}m_{\pi}\bar{\psi}_j\psi_j\vec{a}_0
\label{acoup},
\end{eqnarray}
where $g_{a_0 qq}$ is the coupling constant for $a_0$-quark.
According to these interactions, the amplitude of $t$-channel can
be written as
\begin{eqnarray} \label{tc}
\mathcal{M}_t=g_{a_0\pi \eta }m_\pi\langle N_f |H_{a_0}
|N_i\rangle\frac{1}{t^2-m^2_{a_0}}.
\end{eqnarray}
where  $m_{a_0}$ is the mass of $a_0$.

In the quark model, the nonrelativistic form of Eq. (\ref{coup})
is written as \cite{Li:1997gda,qk3}
\begin{eqnarray}\label{ccpk}
H^{nr}_{m}&=&\sum_j\Big\{\frac{\omega_m}{E_f+M_f}\vsig_j\cdot
\textbf{P}_f+ \frac{\omega_m}{E_i+M_i}\vsig_j \cdot
\textbf{P}_i \nonumber\\
&&-\vsig_j \cdot \textbf{q} +\frac{\omega_m}{2\mu_q}\vsig_j\cdot
\textbf{p}_j\Big\}\frac{I_j}{g_A} \varphi_m,
\end{eqnarray}
and the nonrelativistic form of (\ref{acoup}) is given by
\begin{eqnarray}\label{ccpa}
H^{nr}_{a_0}&=&\sum_j \ m_\pi\left(1+\vsig_j\cdot \textbf{G}_f
\vsig_j\cdot\textbf{G}_i \right) I_j \varphi_m,
\end{eqnarray}
where
\begin{eqnarray}
\textbf{G}_f&=&\mathcal{K}_f\textbf{P}_f+\frac{1}{2m_q}\textbf{p}_j,\\
\textbf{G}_i&=&\mathcal{K}_i\textbf{P}_i+\frac{1}{2m_q}\textbf{p}_j.
\end{eqnarray}
with
\begin{eqnarray}
\mathcal{K}_f\equiv\frac{1}{E_f+M_f},\
\mathcal{K}_i\equiv\frac{1}{E_i+M_i}.
\end{eqnarray}
For emitting a meson, we have $\varphi_m=e^{-i\textbf{q}\cdot
\textbf{r}_j}$, and for absorbing a meson we have
$\varphi_m=e^{i\textbf{q}\cdot \textbf{r}_j}$. In the above
nonrelativistic expansions, vectors $\textbf{r}_{j}$ and
$\textbf{p}_j$ are the internal coordinate and momentum for the
$j$-th quark in the nucleon rest frame. $\omega_m$ and
$\textbf{q}$ are the energy and three-vector momentum of the
meson, respectively. The isospin operator $I_j$ in Eqs.
(\ref{ccpk}) and (\ref{ccpa}) is expressed as
\begin{eqnarray}
I_j=\cases{a^{\dagger}_j(d)a_j(u) & for $\pi^-$\cr 1 & for
$\eta$\cr
\frac{1}{\sqrt{2}}[a^{\dagger}_j(u)a_j(u)-a^{\dagger}_j(d)a_j(d)]
& for $a_0$},
\end{eqnarray}
where $a^{\dagger}_j(d)$ and $a_j(u)$ are the creation and
annihilation operators for the $u$ and $d$ quarks. The axial
vector coupling, $g_A$, relating the hadron spin operator $\vsig$
to the quark spin operator $\vsig_j$ for the $j$-th quark, is
defined as
\begin{eqnarray}
\langle N_f |\sum_j I_j \vsig_j|N_i\rangle\equiv g_A\langle N_f
|\vsig|N_i\rangle,
\end{eqnarray}
which can be explicitly calculated in the NRCQM. For example, for
$\pi^- pn$ vertex, one has $g_A^{\pi^-}=5/3$ and for $\eta NN$,
$g_A^\eta=1$~\cite{Li:1997gda}. The axial vector coupling can then
be related to the $\pi NN$ and $\eta NN$ couplings via the
Goldberger-Treiman relation~\cite{goldberger-treiman}.

\section{amplitudes in quark model}\label{am}

In the calculations, we select the center-mass (c.m.) motion
system for the precess $\pi N\rightarrow \eta N$. The energies and
momenta of the initial meson and nucleon are denoted by
$(\omega_i, \textbf{k})$ and $(E_i, \textbf{P}_i$), while those of
the final state meson and nucleon are denoted by $(\omega_f,
\textbf{q})$ and $(E_f, \textbf{P}_f)$. Note that
$\textbf{P}_i=-\textbf{k}$ and $\textbf{P}_f=-\textbf{q}$.

\subsection{amplitudes for $t$-channel}\label{t}
According to Eq. (\ref{ccpa}), the nonrelativistic scalar coupling
of $H_{a_0}$ for $t$-channel in the c.m. motion system is obtained
as
\begin{eqnarray}\label{tcc}
H^{nr}_{a_0}&=&\sum_j m_\pi\left(1+\vsig_j\cdot \textbf{G}'_f
\vsig_j\cdot\textbf{G}'_i \right) {I^{a_0}_j}
e^{-i(\textbf{q}-\textbf{k})\cdot \textbf{r}_j},
\end{eqnarray}
with
\begin{eqnarray}
\textbf{G}'_f&=&-\mathcal{K}_f\textbf{q}+\frac{1}{2m_q}\textbf{p}_j,\\
\textbf{G}'_i&=&-\mathcal{K}_i\textbf{k}+\frac{1}{2m_q}\textbf{p}_j.
\end{eqnarray}

Substituting Eq. (\ref{tcc}) into Eq. (\ref{tc}), finally we get
the $t$-channel amplitude at quark level, which is given by
\begin{eqnarray}
\mathcal{M}_t&=&g_{a_0\pi \eta }m_\pi\frac{1}{t-m^2_{a_0}}\langle
N_f|\{C_0+C_1+C_2 \textbf{k}\cdot \textbf{q}\nonumber\\
&& +C_3 i\vsig_3\cdot (\textbf{q}\times \textbf{k})\}{3I^{a_0}_3}
e^{-(\textbf{q}-\textbf{k})^2/6\alpha^2}|N_i\rangle,
\end{eqnarray}
where
\begin{eqnarray}
C_0&=&m_\pi\left(1-\frac{1}{2m_q}\frac{1}{2m_q}\frac{\alpha^2}{3}\right)\\
C_3&=&-m_\pi\left[\mathcal{K}_i\mathcal{K}_f+\frac{1}{6m_q}\left(\mathcal{K}_i
+\mathcal{K}_f\right)\right],\\
C_2&=&-m_\pi\left[\mathcal{K}_i\mathcal{K}_f+\frac{1}{6m_q}\left(\mathcal{K}_i
+\mathcal{K}_f+\frac{1}{3m_q}\right)\right]\\
C_1&=&\frac{m_\pi}{6m_q}\left[
\left(\frac{1}{6m_q}+\mathcal{K}_f\right)
\textbf{q}^2+\left(\frac{1}{6m_q}+\mathcal{K}_i
\right)\textbf{k}^2\right].
\end{eqnarray}

To derive the amplitudes for a particular reaction, we have to
transform the amplitudes at quark level into the more familiar
amplitudes at hadronic level, which is given by
\begin{eqnarray}
\mathcal{M}_t&=&g_{a_0\pi \eta }g_{a_0 NN}\frac{m_\pi
}{t-m^2_{a_0}}M_0\{g_1[C_0+C_1+C_2 \textbf{k}\cdot
\textbf{q}]\nonumber\\
&&+g_2C_3 i\vsig\cdot (\textbf{q}\times \textbf{k})\}
e^{-(\textbf{q}-\textbf{k})^2/6\alpha^2},
\end{eqnarray}
with $ g_{1}\equiv\langle N_f |\sum^3_{j=1} I^{a_0}_j|N_i \rangle$
and $ g_{2}\equiv\langle N_f |\sum^3_{j=1} I^{a_0}_j\vsig_j|N_i
\rangle$. In this paper, the coupling constant $g_{\pi \eta a_0}
g_{a_0 NN}$ is obtained from Refs. \cite{aaaa,aa}.

\subsection{amplitudes for $s$-channel}\label{s}

From Eq. (\ref{ccpk}), we obtain nonrelativistic couplings of
$H_{\pi}$ and $H_{\eta}$ for the $s$-channel in the c.m. motion
system, which are written as
\begin{eqnarray}
H_{\pi}&=&\sum_j\frac{I_j}{g^{\pi}_A}\vsig_j \cdot\left[
\textbf{A}_{\pi}e^{i\textbf{k}\cdot
\textbf{r}_j}+\frac{\omega_{\pi}}{2m_q}
\{\textbf{p}_j,e^{i\textbf{k}\cdot
\textbf{r}_j}\}\right] ,\label{hs}\\
H_{\eta}&=&\sum_j \frac{I_j }{g^{\eta}_A}\vsig_j\cdot\left[
\textbf{A}_{\eta}e^{-i\textbf{q}\cdot
\textbf{r}_j}+\frac{\omega_{\eta}}{2m_q}
\{\textbf{p}_j,e^{-i\textbf{q}\cdot \textbf{r}_j}\}\right],
\label{hss}
\end{eqnarray}
with
\begin{eqnarray}
\textbf{A}_{\pi}=-\left(\frac{\omega_\pi}{E_i+M_i}+1\right)\textbf{k},\\
\textbf{A}_{\eta}=-\left(\frac{\omega_\eta}{E_f+M_f}+1\right)\textbf{q}.
\end{eqnarray}

Substituting  Eqs. (\ref{hs}) and (\ref{hss}) into Eq.(\ref{sc}),
then following the procedures used in refs.
\cite{qkk,Li:1997gda,qk3}, we obtain the $s$-channel amplitude in
the harmonic oscillator basis, which is expressed as
\begin{eqnarray}
\mathcal{M}^{s}=\sum_n
(\mathcal{M}^{s}_{3}+\mathcal{M}^{s}_{2})e^{-(\textbf{k}^2+\textbf{q}^2)/6\alpha^2},
\end{eqnarray}
where $\alpha$ is the oscillator strength, and
$e^{-(\textbf{k}^2+\textbf{q}^2)/6\alpha^2}$ is a form factor in
the harmonic oscillator basis. $\mathcal{M}^{s}_{3}$
($\mathcal{M}^{s}_{2}$) corresponds to the amplitudes for the
outgoing meson and incoming meson absorbed and emitted by the same
quark (different quarks). They are given by
\begin{eqnarray}\label{ms3}
\mathcal{M}^{s}_{3}&=&\langle N_f
|\frac{3I_3}{g^{\pi}_A}\Big\{\vsig_{3}\cdot
\textbf{A}_{\eta}\vsig_3 \cdot \textbf{A}_{\pi}\sum_{n=0}
\frac{F_s(n)}{n !}\mathcal{X}^n +\Big[- \vsig_{3}\cdot
\textbf{A}_{\eta}\frac{\omega_{\pi}}{3m_q}\vsig_3 \cdot
\textbf{q}- \frac{\omega_{\eta}}{3m_q}\vsig_3 \cdot
\textbf{k}\vsig_{3}\cdot \textbf{A}_{\pi}+
\frac{\omega_{\eta}}{m_q}\frac{\omega_{\pi}}{m_q}
\frac{\alpha^2}{3}\Big]\nonumber\\
&&\times\sum_{n=1} \frac{F_s(n)}{(n-1) !}\mathcal{X}^{n-1}+
\frac{\omega_{\eta}}{3m_q}\frac{\omega_{\pi}}{3m_q}\vsig_3 \cdot
\textbf{q}\vsig_{3}\cdot \textbf{k}\sum_{n=2} \frac{F_s(n)}{(n-2)
!}\mathcal{X}^{n-2}\Big\}|N_i\rangle,
\end{eqnarray}
and
\begin{eqnarray}\label{ms2}
\mathcal{M}^{s}_{2}&=&\langle N_f
|\frac{6I_1}{g^{\pi}_A}\Big\{\vsig_{1}\cdot
\textbf{A}_{\eta}\vsig_3 \cdot \textbf{A}_{\pi}\sum_{n=0}
\frac{F_s(n)}{n !}\frac{\mathcal{X}^n}{(-2)^n} +\Big[-
\vsig_{1}\cdot \textbf{A}_{\eta}\frac{\omega_{\pi}}{3m_q}\vsig_3
\cdot \textbf{q}-\frac{\omega_{\eta}}{3m_q}\vsig_1 \cdot
\textbf{k}\vsig_{3}\cdot \textbf{A}_{\pi} +
\frac{\omega_{\eta}}{m_q}\frac{\omega_{\pi}}{m_q}
\frac{\alpha^2}{3}\vsig_1\cdot\vsig_3\Big]\nonumber\\
&&\times\sum_{n=1} \frac{F_s(n)}{(n-1)
!}\frac{\mathcal{X}^{n-1}}{(-2)^n}+
\frac{\omega_{\eta}}{3m_q}\frac{\omega_{\pi}}{3m_q}\vsig_1 \cdot
\textbf{q}\vsig_{3}\cdot \textbf{k}\sum_{n=2} \frac{F_s(n)}{(n-2)
!}\frac{\mathcal{X}^{n-2}}{(-2)^{n}}\Big\}|N_i\rangle,
\end{eqnarray}
where $\mathcal{X}\equiv\frac{\textbf{k}\cdot \textbf{q}}{3
\alpha^2}$. The subscriptions of the spin operator $\vsig$ denote
that it either operates on quark 3 or quark 1.

In the Eqs. (\ref{ms3}) and (\ref{ms2}), the factor $F_s(n)$ is
given by expanding the energy propagator in Eq. (\ref{sc}) (and
similarly in Eq. (\ref{uc})) which leads to
\begin{eqnarray}
F_s(n)=\frac{M_n}{P_i \cdot k-nM_n \omega_h}
\end{eqnarray}
where $n$ is the total excitation quantum number in the harmonic
oscillator basis; $M_n$ is the mass of the excited state in the
$n$-th shell, while $\omega_h$ is the typical energy of the
harmonic oscillator; $P_i$ and $k$ are the four momenta of the
initial state nucleons and incoming $\pi^-$ mesons in the c.m.
system. This factor has clear physical meaning that recovers the
hadronic level propagators. We will come back to this in the next
section.

The above two transitions can be written coherently in terms of a
number of $g$-factors, which will allow us to relate the
quark-level amplitudes to those at hadronic level
\begin{eqnarray} \label{sac}
\mathcal{M}^{s}&=&\frac{1}{g^{\pi}_A}\Big\{\textbf{A}_{\eta}\cdot\textbf{A}_{\pi}\sum_{n=0}
\left[g_{s1}+(-2)^{-n}g_{s2}\right] \frac{F_s(n)}{n
!}\mathcal{X}^n
+\left(-\frac{\omega_{\pi}}{3m_q}\textbf{A}_{\eta}\cdot
\textbf{q}-\frac{\omega_{\eta}}{3m_q}\textbf{A}_{\pi}\cdot
\textbf{k}+\frac{\omega_{\eta}}{m_q}\frac{\omega_{\pi}}{m_q}
\frac{\alpha^2}{3}\right)\nonumber\\
&&\times\sum_{n=1}[g_{s1}+(-2)^{-n}g_{s2}] \frac{F_s(n)}{(n-1)
!}\mathcal{X}^{n-1}+
\frac{\omega_{\eta}\omega_{\pi}}{(3m_q)^2}\textbf{k}\cdot\textbf{q}
\sum_{n=2}\frac{F_s(n)}{(n-2)!}[g_{s1}+(-2)^{-n}g_{s2}]\mathcal{X}^{n-2}\nonumber\\
&&+i\vsig
\cdot(\textbf{A}_{\eta}\times\textbf{A}_{\pi})\sum_{n=0}\left[g_{v1}+(-2)^{-n}g_{v2}\right]
\frac{F_s(n)}{n !}\mathcal{X}^n
+\frac{\omega_{\eta}\omega_{\pi}}{(3m_q)^2}i\vsig\cdot(\textbf{q}\times\textbf{k})
\nonumber\\
&&
\times\sum_{n=2}\left[g_{v1}+(-2)^{-n}g_{v2}\right]\frac{F_s(n)}{(n-2)!}\mathcal{X}^{n-2}\Big\}e^{-(\textbf{k}^2+\textbf{q}^2)/6\alpha^2}.
\end{eqnarray}
where the $g$ factors are defined as
\begin{eqnarray}
g_{s1}&\equiv&\langle N_f |\sum^3_{j=1} I_j|N_i \rangle,\\
g_{v1}&\equiv&\langle N_f |\sum^3_{j=1} I_j\sigma_{jz}|N_i \rangle,\\
g_{s2}&\equiv&\langle N_f |\sum_{i\neq j} I_j\vsig_i\cdot \vsig_j
|N_i
\rangle/3,\\
g_{v2}&\equiv&\langle N_f |\sum_{i\neq j} I_j(\vsig_i\times
\vsig_j)_z |N_i\rangle/2.
\end{eqnarray}
The numerical values of these $g$ factors can be derived in the
$SU(6)\otimes O(3)$ symmetry limit.

\subsection{amplitudes for $u$-channel}

According to Eq.~(\ref{ccpk}), the nonrelativistic expansions of
the $u$-channel meson-nucleon interactions can also be derived
\begin{eqnarray}
H_{\pi}&=&\sum_j \frac{I_j}{g^{\pi}_A}\vsig_j \cdot\left[
\textbf{B}_{\pi}e^{i\textbf{k}\cdot
\textbf{r}_j}+\frac{\omega_{\pi}}{2m_q}
\{\textbf{p}_j,e^{i\textbf{k}\cdot \textbf{r}_j}\}\right],\\
H_{\eta}&=&\sum_j\frac{I_j}{g^{\eta}_A}\vsig_j \cdot\left[
\textbf{B}_{\eta}e^{-i\textbf{q}\cdot
\textbf{r}_j}+\frac{\omega_{\eta}}{2m_q}\{
\textbf{p}_j,e^{-i\textbf{q}\cdot \textbf{r}_j}\}\right],
\end{eqnarray}
where
\begin{eqnarray}
\textbf{B}_{\pi}&=&-\omega_\pi\left(\mathcal{K}_f+\mathcal{K}_j\right)\textbf{q}
-\left(\omega_\pi\mathcal{K}_j+1\right)\textbf{k},\\
\textbf{B}_{\eta}&=&-\omega_\eta\left(\mathcal{K}_i+\mathcal{K}_j\right)\textbf{k}
-\left(\omega_\eta\mathcal{K}_i+1\right)\textbf{q},
\end{eqnarray}
with $\mathcal{K}_j\equiv1/(E_j+M_j)$.

Following the same procedure in \ref{s}, we obtain amplitudes for
the outgoing meson and incoming meson absorbed and emitted by the
same quark
\begin{eqnarray}\label{mu3}
\mathcal{M}^{u}_{3}&=&-\langle N_f
|\frac{3I_3}{g^{\pi}_A}\Big\{\vsig_{3}\cdot
\textbf{B}_{\pi}\vsig_3 \cdot \textbf{B}_{\eta}\sum_{n=0}
F_u(n)\frac{1}{n !}\mathcal{X}^n + \Big[-\vsig_{3}\cdot
\textbf{B}_{\pi}\frac{\omega_{\eta}}{3m_q}\vsig_3 \cdot \textbf{k}
- \frac{\omega_{\pi}}{3m_q}\vsig_3 \cdot \textbf{q}\vsig_{3}\cdot
\textbf{B}_{\eta}+
\frac{\omega_{\eta}}{m_q}\frac{\omega_{\pi}}{m_q}
\frac{\alpha^2}{3}\Big]\nonumber\\
&& \times\sum_{n=1} F_u(n)\frac{\mathcal{X}^{n-1}}{(n-1) !}+
\frac{\omega_{\eta}}{3m_q}\frac{\omega_{\pi}}{3m_q}\vsig_3 \cdot
\textbf{k}\vsig_{3}\cdot \textbf{q}\sum_{n=2}
F_u(n)\frac{\mathcal{X}^{n-2}}{(n-2) !}\Big\}|N_i\rangle,
\end{eqnarray}
and by different quarks
\begin{eqnarray}\label{mu2}
\mathcal{M}^{u}_{2}&=&-\langle N_f
|\frac{6I_1}{g^{\pi}_A}\Big\{\vsig_{1}\cdot
\textbf{B}_{\pi}\vsig_3 \cdot \textbf{B}_{\eta}\sum_{n=0}
\frac{F_u(n)}{n !}\frac{\mathcal{X}^n }{(-2)^n}+\Big[
-\vsig_{1}\cdot \textbf{B}_{\pi}\frac{\omega_{\eta}}{3m_q}\vsig_3
\cdot \textbf{k}-\frac{\omega_{\pi}}{3m_q}\vsig_1 \cdot
\textbf{q}\vsig_{3}\cdot \textbf{B}_{\eta}+
\frac{\omega_{\eta}}{m_q} \frac{\omega_{\pi}}{m_q}
\frac{\alpha^2}{3}\vsig_1 \cdot \vsig_3\Big]\nonumber\\&&\times
\sum_{n=1} \frac{F_u(n)}{(n-1) !}\frac{\mathcal{X}^{n-1}}{(-2)^n}+
\frac{\omega_{\eta}}{3m_q}\frac{\omega_{\pi}}{3m_q}\vsig_1 \cdot
\textbf{k}\vsig_{3}\cdot \textbf{q}\sum_{n=2} \frac{F_u(n)}{(n-2)
!}\frac{\mathcal{X}^{n-2}}{(-2)^{n}}\Big\}|N_i\rangle ,
\end{eqnarray}
where the factor $F_u(n)$, which can be related to the
propagators, is written as
\begin{eqnarray}
F_u(n)=\frac{M_n}{P_i \cdot q+nM_n \omega_h},
\end{eqnarray}
where $q$ are the four momenta of the outgoing $\eta$ mesons in
the c.m. system.

The total amplitude for the $u$-channel is expressed as
\begin{eqnarray}\label{uac}
\mathcal{M}^{u}&=&\frac{-1}{g^{\pi}_A}\Big\{\textbf{B}_{\pi}\cdot\textbf{B}_{\eta}\sum_{n=0}
\left[g_{s1}+(-2)^{-n}g_{s2}\right] \frac{F_u(n)}{n
!}\mathcal{X}^n
+\left(-\frac{\omega_{\eta}}{3m_q}\textbf{B}_{\pi}\cdot
\textbf{k}-\frac{\omega_{\pi}}{3m_q}\textbf{B}_{\eta}\cdot
\textbf{q}+\frac{\omega_{\pi}}{m_q}\frac{\omega_{\eta}}{m_q}
\frac{\alpha^2}{3}\right)\nonumber\\
&&\times\sum_{n=1}[g_{s1}+(-2)^{-n}g_{s2}] \frac{F_u(n)}{(n-1)
!}\mathcal{X}^{n-1}+
\frac{\omega_{\eta}\omega_{\pi}}{(3m_q)^2}\textbf{k}\cdot\textbf{q}
\sum_{n=2}\frac{F_u(n)}{(n-2)!}[g_{s1}+(-2)^{-n}g_{s2}]
\mathcal{X}^{n-2} \nonumber\\
&& + i\vsig
\cdot(\textbf{B}_{\pi}\times\textbf{B}_{\eta})\sum_{n=0}\left[g_{v1}+(-2)^{-n}g_{v2}\right]
\frac{F_u(n)}{n !}\mathcal{X}^n
-\frac{\omega_{\eta}\omega_{\pi}}{(3m_q)^2}i\vsig\cdot(\textbf{q}\times\textbf{k})
\sum_{n=2}[g_{v1}+(-2)^{-n}g_{v2}]\nonumber\\
&&\times\frac{F_u(n)}{(n-2)!}\mathcal{X}^{n-2} +i\vsig \cdot
\left[-\frac{\omega_{\eta}}{3m_q}(\textbf{B}_{\pi}\times
\textbf{k})-\frac{\omega_{\pi}}{3m_q}(\textbf{q}\times\textbf{B}_{\eta}
)\right ]\sum_{n=1}\left[g_{v1}+(-2)^{-n}g_{v2}\right]
\mathcal{X}^{n-1}
\frac{F_u(n)}{(n-1) !} \Big\}\nonumber\\
&&\times e^{-(\textbf{k}^2+\textbf{q}^2)/6\alpha^2}.
\end{eqnarray}

The first terms in Eqs. (\ref{ms3}), (\ref{ms2}), (\ref{mu3}) and
(\ref{mu2}) come from the correlation between the c.m. motion of
the pion meson transition operator and the c.m. motion of
$\eta$-meson transition operator; the second and the third terms
are the correlation among the internal and the c.m. motions of the
$\pi^-$ and $\eta$ transition operators, and their contributions
begin with the $n\geq 1$ exited states in the harmonic oscillator
basis. The last two terms in these equations correspond to the
correlation of the internal motion between the $\pi^-$ and $\eta$
transition operators, and their contributions begin with either
$n\geq 1$ or $n\geq 2$ exited states. The higher shell resonance
amplitudes are suppressed remarkably by the factors $1/n!$ and $
\mathcal{X}^n\equiv\left(\textbf{k}\cdot \textbf{q}/3
\alpha^2\right)^n$, which come from the spacial integral.

\section{separation of the resonance contributions} \label{rs}

The obtained amplitudes, $\mathcal{M}^s$ and $\mathcal{M}^u$,
involve excited states with the total excitation quantum number
$n$ in the harmonic oscillator basis, which are degenerate to each
other. To see the contributions of individual resonances, we need
to separate out the single resonance excitation amplitudes for
each $n$. In this work we only separate out the resonance
excitation amplitudes for the $s$-channel, and treat the
resonances in the $u$-channel as degenerate to $n$. This is
because the resonances in the $u$-channel contribute virtually and
are generally suppressed by the kinematics.

In the amplitudes for the $s$- and $u$-channels, the factors
$F_s(n)$ and $F_u(n)$ can be rewritten as
\begin{eqnarray}
F_s(n)=\frac{2M_n}{s-M^2_n},\label{bws}\\
F_u(n)=\frac{-2M_n}{u-M^2_n},
\end{eqnarray}
where $s$ $[=(P_i+k)^2]$ and $u$ $[=(P_i-q)^2]$ are the Mandelstam
variables. Taking into account the effects of the resonance mass
and width, we thus substitute a Breit-Wigner distribution for
$F_s(n)$, i.e.
\begin{eqnarray}\label{wd}
F_s(n)\rightarrow F_s(R)= \frac{2M_R}{s-M^2_R+iM_R \Gamma_R},
\end{eqnarray}
where $M_R$ and $\Gamma_R$ are the resonance mass and width,
respectively. The resonance transition amplitudes in the $s$ and
$u$-channels can be generally expressed as
\begin{eqnarray}
\mathcal{M}^s_R=\frac{2M_R}{s-M^2_R+iM_R
\Gamma_R}\mathcal{O}_Re^{-(\textbf{k}^2+\textbf{q}^2)/6\alpha^2},
\end{eqnarray}
and
\begin{eqnarray}
\mathcal{M}^u_n=\frac{2M_n}{u-M^2_n}\mathcal{O}_n
e^{-(\textbf{k}^2+\textbf{q}^2)/6\alpha^2},
\end{eqnarray}
respectively, where $\mathcal{O}_R$ and $\mathcal{O}_n$ are
determined by the structure of each resonance and their couplings
to the meson and nucleon.

It should be pointed out that the introduction of the Breit-Wigner
widths in the $s$-channel is arbitrary in this framework, where
the width effects from intermediate resonances cannot be
automatically produced. However, since it allows a separation of
individual resonances, the inclusion of resonance widths from the
experiments will make an explicit connection between the
transition amplitudes and individual resonance contributions. It
should also be mentioned that such an analytic advantage will only
appear in the NRQCD model where a harmonic oscillator potential is
employed.

\subsection{n=0 shell resonances}

 For $n=0$, only the nucleon pole term contributes to
the transition amplitude. Its $s$-channel amplitude is
\begin{eqnarray}
\mathcal{M}_N^{s}&=&\mathcal{O}_N
\frac{2M_0}{s-M^2_0}e^{-(\textbf{k}^2+\textbf{q}^2)/6\alpha^2},
\end{eqnarray}
with
\begin{eqnarray}
\mathcal{O}_N=\left[g_{s1}+g_{s2}\right]\textbf{A}_{\eta}\cdot\textbf{A}_{\pi}
+\left[g_{v1}+g_{v2}\right]i\vsig
\cdot(\textbf{A}_{\eta}\times\textbf{A}_{\pi}),\nonumber\\
\end{eqnarray}
where $M_0$ is the nucleon mass.

\subsection{n=1 shell resonances}\label{n1}
For $n=1$, only $S$- and $D$-waves contribute in the $s$-channel.
Note that the spin independent amplitude for $D$-waves is
proportional to the Legendre function $P^0_2(\cos\theta)$, and the
spin dependent amplitude for $D$-waves is in proportion to
$\frac{\partial}{\partial \theta}P^0_2(\cos\theta)$. Moreover, the
$S$-wave amplitude is independent of the scattering angle. Thus,
the $S$- and $D$-wave amplitudes can be separated out easily. They
are presented as
\begin{eqnarray}
\mathcal{M}^{s}(S)&=&\mathcal{O}_S F_s(R)e^{-(\textbf{k}^2+\textbf{q}^2)/6\alpha^2},\\
\mathcal{
M}^{s}(D)&=&\mathcal{O}_DF_s(R)e^{-(\textbf{k}^2+\textbf{q}^2)/6\alpha^2},
\end{eqnarray}
with
\begin{eqnarray}
\mathcal{O}_S&=&\left(g_{s1}-\frac{1}{2}g_{s2}\right)
\Big(|\textbf{A}_{\eta}||\textbf{A}_{\pi}|
\frac{|\textbf{k}||\textbf{q}|}{9
\alpha^2}-\frac{\omega_{\pi}}{3m_q}\textbf{A}'_{\eta}\cdot
\textbf{q}\nonumber\\
&&-\frac{\omega_{\eta}}{3m_q}\textbf{A}_{\pi}\cdot
\textbf{k}+\frac{\omega_{\eta}}{m_q}\frac{\omega_{\pi}}{m_q}
\frac{\alpha^2}{3}\Big),\\
\mathcal{O}_D&=&\left(g_{s1}-\frac{1}{2}g_{s2}\right)|\textbf{A}_{\eta}||\textbf{A}_{\pi}|
(3\cos^2 \theta-1)\frac{|\textbf{k}||\textbf{q}|}{9 \alpha^2}\nonumber\\
&&+\left(g_{v1}-\frac{1}{2}g_{v2}\right)i \vsig
\cdot(\textbf{A}_{\eta}\times\textbf{A}_{\pi})
\frac{\textbf{k}\cdot \textbf{q}}{3 \alpha^2}.
\end{eqnarray}

For the $S$-waves, the possible resonances are $S_{11}(1535)$
($[\textbf{70},^2\textbf{8}]$) and $S_{11}(1650)$
($[\textbf{70},^4\textbf{8}]$); and for the $D$-waves, the
possible resonances are $D_{13}(1520)$
($[\textbf{70},^2\textbf{8}]$), $D_{13}(1700)$
($[\textbf{70},^4\textbf{8}]$)  and $D_{15}(1675)$
($[\textbf{70},^4\textbf{8}]$). The separated amplitudes for the
$S$ and $D$-wave can thus be re-written as
\begin{eqnarray}
\mathcal{M}^{s}(S)&=&[g_{S_{11}(1535)}+g_{S_{11}(1650)}]\mathcal{M}^{s}(S),\\
 \mathcal{M}^{s}(D)&=&[g_{D_{13}(1520)}+g_{D_{13}(1700)}+g_{D_{15}(1675)}]\nonumber\\
&&\times\mathcal{M}^{s}(D),
\end{eqnarray}
where the factor $g_R$ ($R=S_{11}(1535)$, etc) represents the
resonance transition strengths in the spin-flavor space, and is
determined by the matrix elements $\langle
N_f|H_\eta|N_j\rangle\langle N_j|H_\pi|N_i\rangle$. Their relative
strengths can be explicitly determined by the following relations
\begin{eqnarray}
\frac{g_{S_{11}(1535)}}{g_{S_{11}(1650)}}=\frac{\langle
N_f|\vsig_3|S_{11}(1535)\rangle\langle
S_{11}(1535)|I_3\vsig_3|N_i\rangle}{\langle
N_f|\vsig_3|S_{11}(1650)\rangle\langle
S_{11}(1650)|I_3\vsig_3|N_i\rangle},\\
\frac{g_{D_{13}(1520)}}{g_{D_{13}(1700)}}=\frac{\langle
N_f|\vsig_3|D_{13}(1520)\rangle\langle
D_{13}(1520)|I_3\vsig_3|N_i\rangle}{\langle
N_f|\vsig_3|D_{13}(1700)\rangle\langle
D_{13}(1700)|I_3\vsig_3|N_i\rangle},\\
\frac{g_{D_{13}(1700)}}{g_{D_{15}(1675)}}=\frac{\langle
N_f|\vsig_3|D_{13}(1700)\rangle\langle
D_{13}(1700)|I_3\vsig_3|N_i\rangle}{\langle
N_f|\vsig_3|D_{15}(1675)\rangle\langle
D_{15}(1675)|I_3\vsig_3|N_i\rangle}.
\end{eqnarray}
The determined values are listed in Tab. \ref{factor}.

Finally, we obtain the partial amplitudes for individual
resonances
\begin{eqnarray}
\mathcal{M}^s(S_{11}(1535))&=&g_{S_{11}(1535)}\mathcal{M}^{s}(S),\\
\mathcal{M}^s(S_{11}(1650))&=&g_{S_{11}(1650)}\mathcal{M}^{s}(S),\\
\mathcal{
M}^{s}(D_{13}(1520))&=&g_{D_{13}(1520)}\mathcal{M}^{s}(D),\\
 \mathcal{M}^{s}(D_{13}(1700))&=&g_{D_{13}(1700)}\mathcal{M}^{s}(D),\\
 \mathcal{M}^{s}(D_{15}(1675))&=&g_{D_{15}(1675)}
 \mathcal{M}^{s}(D).
\end{eqnarray}

\subsection{n=2 shell resonances}
For $n=2$, only the $P$ and $F$-wave are involved in the
$s$-channel. Note that the spin-independent amplitude for the
$P$-wave is in proportion to $P^0_1(\cos\theta)$, and the
spin-dependent amplitude for the $P$-wave is in proportion to
$\frac{\partial}{\partial \theta}P^0_1(\cos\theta)$; the
spin-independent amplitude for the $F$-wave is in proportion to
$P^0_3(\cos\theta)$, and the spin-dependent amplitude for the
$F$-wave is in proportion to $\frac{\partial}{\partial
\theta}P^0_3(\cos\theta)$. Thus, the $P$- and $F$-wave amplitudes
can be separated out. They are given by
\begin{eqnarray}
\mathcal{M}^{s}(P)&=&\mathcal{O}_P F_s(R)e^{-(\textbf{k}^2+\textbf{q}^2)/6\alpha^2},\\
\mathcal{
M}^{s}(F)&=&\mathcal{O}_FF_s(R)e^{-(\textbf{k}^2+\textbf{q}^2)/6\alpha^2},
\end{eqnarray}
with
\begin{eqnarray}
\mathcal{O}_P&=&\Big[\left(g_{s1}+\frac{1}{4}g_{s2}\right)
\Big(|\textbf{A}_{\eta}||\textbf{A}_{\pi}|
\frac{|\textbf{k}||\textbf{q}|}{10
\alpha^2}-\frac{\omega_{\pi}}{3m_q}\textbf{A}_{\eta}\cdot
\textbf{q}-\frac{\omega_{\eta}}{3m_q}\textbf{A}_{\pi}\cdot
\textbf{k}+\frac{\omega_{\eta}}{m_q}\frac{\omega_{\pi}}{m_q}
\frac{\alpha^2}{3}\Big)\nonumber\\
&& +\left(g_{s1}+\frac{1}{4}g_{s2}\right)
\frac{3\alpha^2\omega_{\eta}\omega_{\pi}}{(3m_q)^2}\Big]
\frac{|\textbf{k}||\textbf{q}|}{3 \alpha^2}\cos
\theta+\left(g_{v1}+\frac{1}{4}g_{v2}\right)
\frac{\omega_{\eta}\omega_{\pi}}{(3m_q)^2}i\vsig\cdot(\textbf{q}\times\textbf{k})
\nonumber\\
&&+\frac{1}{10}\left(g_{v1}+\frac{1}{4}g_{v2}\right) i \vsig
\cdot(\textbf{A}_{\eta}\times\textbf{A}_{\pi})\left(\frac{|\textbf{k}||\textbf{q}|}{3 \alpha^2}\right)^2,\\
\mathcal{O}_F&=&\left(g_{s1}+\frac{1}{4}g_{s2}\right)\frac{1}{2}|\textbf{A}_{\eta}||\textbf{A}_{\pi}|
\left(\cos^3 \theta-\frac{3}{5}\cos
\theta\right)\left(\frac{|\textbf{k}||\textbf{q}|}{3
\alpha^2}\right)^2 +\left(g_{v1}+\frac{1}{4}g_{v2}\right)i \vsig
\cdot(\textbf{A}_{\eta}\times\textbf{A}_{\pi})\nonumber\\
&&\times\frac{1}{2}\left(\cos^2 \theta -\frac{1}{5}
\right)\left(\frac{|\textbf{k}||\textbf{q}|}{3 \alpha^2}\right)^2.
\end{eqnarray}
For the $P$-wave, the possible resonances are $P_{11}(1440)$
($[\textbf{56},^2\textbf{8}]$), $P_{13}(1720)$
($[\textbf{56},^2\textbf{8}]$), $P_{11}(1710)$
($[\textbf{70},^2\textbf{8}]$), $P_{13}(1900)$
($[\textbf{70},^4\textbf{8}]$, $[\textbf{70},^2\textbf{8}]$),
$P_{11}(2100)$ ($[\textbf{70},^4\textbf{8}]$); and for the
$F$-wave, the possible resonances are $F_{15}(1680)$
($[\textbf{56},^2\textbf{8}]$), $F_{17}(1990)$
($[\textbf{70},^4\textbf{8}]$) and $F_{15}(2000)$
($[\textbf{70},^2\textbf{8}]$, $[\textbf{70},^4\textbf{8}]$). Thus
the amplitudes for the $P$ and $F$-wave can be re-written as
\begin{eqnarray}
\mathcal{M}^{s}(P)&=&[g_{P_{11}(1440)}+g_{P_{11}(1710)}+g_{P_{13}(1720)}\nonumber\\
&&+g_{P_{13}(1900)}]\mathcal{M}^{s}(S),\label{ad}\\
 \mathcal{M}^{s}(F)&=&[g_{F_{15}(1680)}+g_{F_{15}(2000)}]\mathcal{M}^{s}(D)\label{ad1},
\end{eqnarray}
with the same method applied in \ref{n1}, we can determine the
$g_R$ factors in Eqs. (\ref{ad}) and (\ref{ad1}). The $g$ and
$g_R$ factors given by the quark model are listed in Tab.
\ref{factor}. We find that $g_{D_{13}(1700)}$, $g_{P_{13}(1900)}$
and $g_{P_{11}(2100)}$ is about an order of magnitude less than
those of other resonances. Thus, the contributions of
$D_{13}(1700)$, $P_{13}(1900)$ and $P_{11}(2100)$ are negligible.

The higher resonances (i.e. $n\geq 3$) are treated as degenerate,
for they are less important at the energy region near the
threshold of the $\eta N$ production.

\begin{table}[ht]
\caption{various g and $g_R$ factors in quark model.}
\label{factor}
\begin{tabular}{|c|c|c|c|c|c|c|c|c }\hline\hline
factor & value               & &  factor  &  value        &&  factor          & value\\
\hline
$g_{s1}$ & 1               & &  $g_{S_{11}(1535)}$  &  2        &&  $g_2$          & 5/3 \\
$g_{s2}$ & 2/3              &&  $g_{S_{11}(1650)}$  & -1        &&   $g_{P_{11}(1710)}$  &180/619  \\
$g_{v1}$ & 5/3             & &  $g_{D_{13}(1520)}$  & 2         &&  $g_{P_{13}(1900)}$& 18/619 \\
$g_{v2}$ & 0                &&  $g_{D_{13}(1700)}$  & -1/10     &&  $g_{P_{11}(2100)}$& -16/619 \\
$g^{\pi}_{A}$ & 5/3         &&  $g_{D_{15}(1675)}$  & -9/10     && $g_{F_{15}(1680)}$ &  5/3 \\
$g^{\eta}_{A}$ & 1          &&  $g_{P_{11}(1440)}$  & 225/619   &&  $g_{F_{15}(2000)}$& -2/21     \\
$g_1$             &  1    && $g_{P_{13}(1720)}$   & 180/619   && $g_{F_{17}(1990)}$ & -4/7 \\
\hline
\end{tabular}
\end{table}

\begin{table}[ht]
\caption{Breit-Wigner masses $M_R$ (in MeV) and widths $\Gamma_R$
(in MeV) for the resonances. $n=1$ and $n=2$ stand for the
degenerate states with quantum number $n=1$ and $n=2$ in the
$u$-channel.} \label{parameter}
\begin{tabular}{|c|c|c||c|c|c| }\hline\hline
resonance &\ \  $M_R$ \ \ & \ \ $\Gamma_R$ \ \ &resonance&\ \ $M_R$ \ \ & \ \ $\Gamma_R$ \ \  \\
\hline
$S_{11}(1535)$& 1535    &150 & $P_{11}(1440)$& 1440    &300   \\
$S_{11}(1650)$& 1655    &165 & $P_{11}(1710)$& 1710    &100   \\
$D_{13}(1520)$& 1520    &115 & $P_{13}(1720)$& 1720    &200 \\
$D_{13}(1700)$& 1700    &115 & $P_{13}(1900)$& 1900    &500 \\
$D_{15}(1675)$& 1675    &150 & $P_{11}(2100)$& 2100    &150   \\
n=1           & 1650    &230 & $F_{15}(1680)$& 1685    &130     \\
n=2           & 1750    &300 & $F_{15}(2000)$& 2000    &200   \\
-           & -    &- & $F_{17}(1990)$& 1990    &350   \\
\hline
\end{tabular}
\end{table}

\section{calculations and analysis}\label{RD}

\subsection{parameters}

Since the resonance amplitudes have been obtained, one can
calculate the differential cross section with
\begin{eqnarray}
\frac{d\vsig}{d\Omega}=\frac{(E_i+M_i)(E_f+M_f)}{64\pi^2
s}\frac{|\textbf{q}|}{|\textbf{k}|}\frac{1}{2}
\sum_{\lambda_i,\lambda_f}|\mathcal{M}_{\lambda_f,\lambda_i}|^2 ,
\end{eqnarray}
where $\lambda_i=\pm 1/2$ and $\lambda_f=\pm 1/2$ are the
helicities of the initial and final state nucleons, respectively.

To take into account the relativistic effects, as done in
\cite{qkka} , we introduce the Lorentz boost factor in the spatial
part of the amplitudes, which is
\begin{eqnarray}
\mathcal{O}_i(\textbf{k},\textbf{q})\rightarrow \gamma_k \gamma_q
\mathcal{O}_i(\textbf{k}\gamma_k, \textbf{q} \gamma_q),
\end{eqnarray}
where $\gamma_k=M_i/E_i$ and $\gamma_q=M_f/E_f$.

In the calculations, the quark-pseudoscalar-meson couplings are
the overall parameters in the $s$ and $u$-channel transitions.
However, they are not totally free ones. They can be related to
the hadronic couplings via the Goldberger-Treiman
relation~\cite{goldberger-treiman}:
\begin{equation}
g_{m NN} =\frac{g_A^m M_N}{f_m} \ ,
\end{equation}
where $m$ denotes the pseudoscalar mesons, $\pi$, $\eta$, etc;
$f_m$ is the meson decay constant defined earlier and $g_A^m$ is
the axial vector coupling for the meson.

The $\pi NN$ coupling $g_{\pi NN}$ is a well-determined number:
\begin{eqnarray}
g_{\pi NN}=13.48,
\end{eqnarray}
thus we fix it in our calculations. The $\eta NN$ coupling is the
only free parameter in the present calculations, and to be
determined by the experimental data. This quantity has not been
well established in both experiment and theory. Its values
extracted from different models are still controversial and
possess large uncertainties. By fitting the data (differential
cross section) at $W\leq 1524$ MeV, we find that our calculations
favor a small $\eta NN$ coupling around $g_{\eta NN}=0.81$, which
is comparable with those deduced from fitting the $\eta$
photo-production~\cite{qk4,Li:1998ni,qkka}. The small $\eta NN$
coupling is also predicted in Refs.
\cite{Grein:1979nw,Tiator:1994et,Kirchbach:1996kw,Stoks:1999bz,Piekarewicz,Zhu:2000eh}.
In contrast, the $\eta NN$ coupling derived here is much smaller
than those used/predicted
in~\cite{cc1,cc2,cc3,cc4,Nasrallah:2005hn}, which are in a range
of $g_{\eta NN}=4\sim 9$.  It should be noted that we do not
expect that one parameter fitting can provide an overall
description of the experimental data. Therefore, we only consider
the data at $W\leq 1524$ MeV as a reasonable constraint on the
$g_{\eta NN}$ and calculation results with $W > 1524$ MeV will
present as a prediction.

For the $a_0\pi\eta$ and $a_0 NN$ couplings we adopt a commonly
used value $g_{a_{0}NN}g_{a_0 \pi \eta} =100$ in the
calculation~\cite{aaaa,aa}.

There are other two overall parameters, $m_q$ and $\alpha$, from
the quark model. In the calculation we adopt their standard values
in the the quark model,
\begin{eqnarray}
  m_q&=&330 \ \mathrm{MeV},\\
 \alpha^2&=&0.16 \ \mathrm{GeV}^2.
\end{eqnarray}

For those $s$-channel resonances which generally have a broad
width, the treatment for their widths to be constants is not
appropriate. Thus, we take the final-state-momentum-dependent
width~\cite{qkk,qkk0,qk3,qk4}:
\begin{eqnarray}
\Gamma(\textbf{q})=\Gamma_R\frac{\sqrt{s}}{M_R}\sum_i x_i
\left(\frac{|\textbf{q}_i|}{|\textbf{q}^R_i|}\right)^{2l+1}
\frac{D(\textbf{q}_i)}{D(\textbf{q}^R_i)},
\end{eqnarray}
where $|\textbf{q}^R_i|=((M_R^2-M_0^2+m_i^2)/4M_R^2-m_i^2)^{1/2}$,
and $|\textbf{q}_i|=((s-M_0^2+m_i^2)/4s-m_i^2)^{1/2}$; $x_i$ is
the branching ratio of the resonance decaying into a meson with
mass $m_i$ and a nucleon, and $\Gamma_R$ is the total decay width
of the $s$-channel resonance with mass $M_R$.
$D(\textbf{q})=e^{-\textbf{q}^2/3\alpha^2}$ is a fission barrier
function.

We adopt the PDG values for the resonance masses and
widths~\cite{PDG}, which are listed in Tab. \ref{parameter}. The
contributions of $u$-channel for $n\geq 1$ shells are negligibly
small, which are insensitive to the degenerate masses and widths
for these shells. In this work, we take $M_1=1650$ MeV ($M_2=1750$
MeV), $\Gamma_1=230$ MeV ($\Gamma_2=300$ MeV) for the degenerate
mass and width of $n=1$ ($n=2$) shell, respectively.

\begin{widetext}
\begin{center}
\begin{figure}[ht]
\centering \epsfxsize=8.5 cm \epsfbox{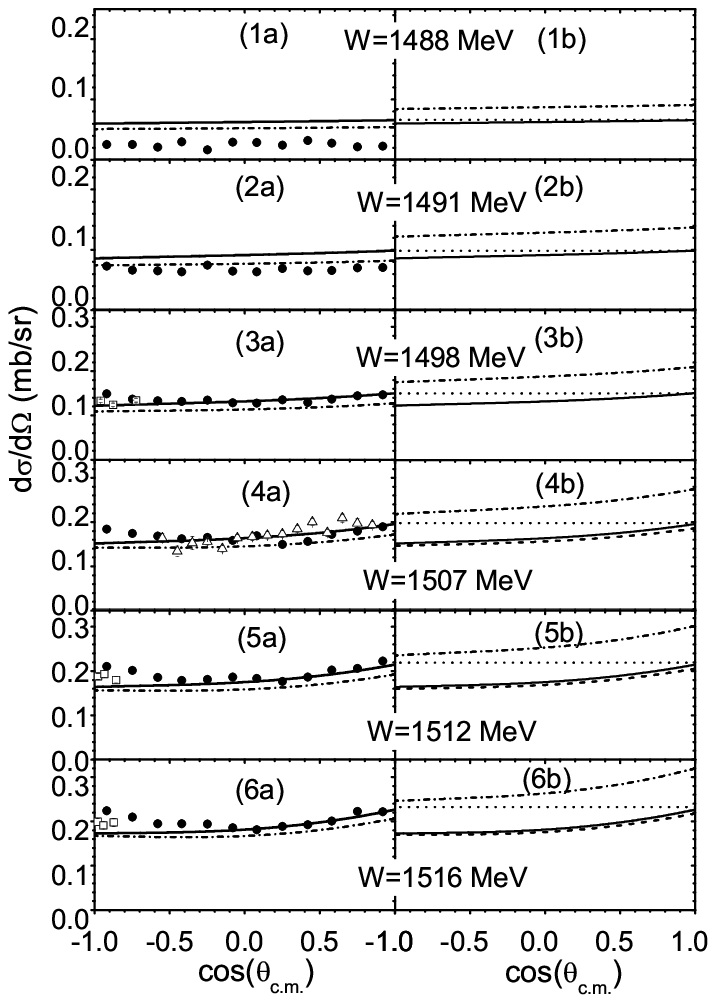} \centering
\epsfxsize=8.5 cm \epsfbox{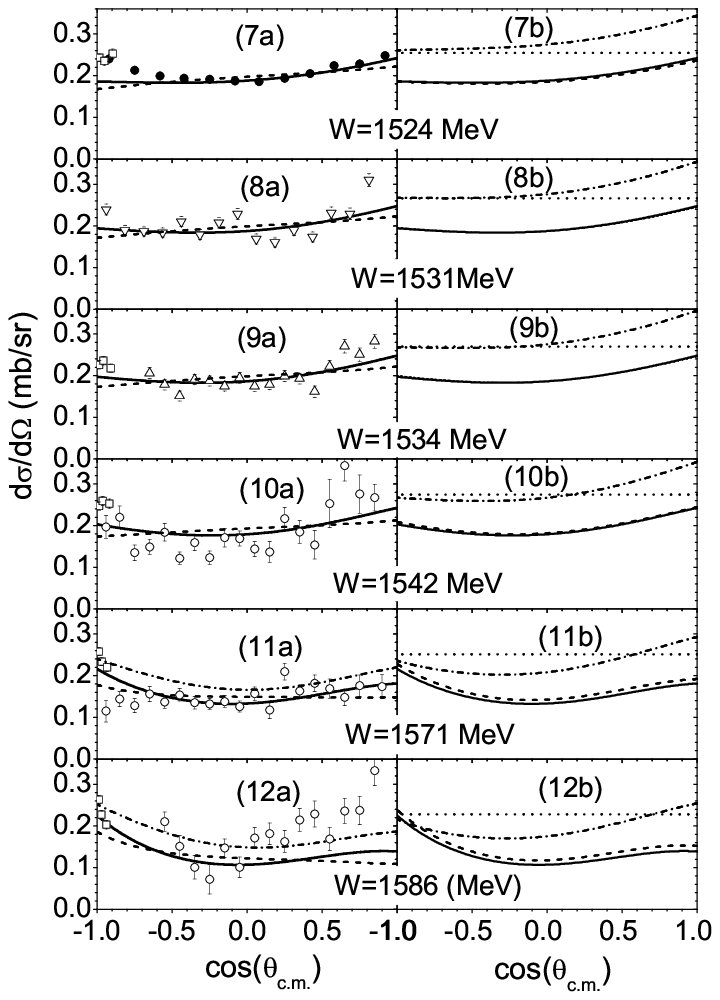} \centering \caption{The
differential cross sections at various $W$. The data are from
\cite{exp1} (open circles), \cite{exp3} (open up-triangles),
\cite{exp4} (open down-triangles), \cite{exp6} (open squares), and
the recent experiment \cite{exp7} (solid circles). The solid
curves are for the full model differential cross sections. In
(1a-12a), the dash-dotted and dashed curves are for the results
switched off the contributions from nucleon pole and
$D_{13}(1520)$, respectively. In (1b-12b), the dash-dotted and
dashed curves correspond to the results without $S_{11}(1650)$ and
without $t$-channel, respectively; the straight lines corresponds
to the partial differential cross sections for $S_{11}(1535)$.
}\label{fig-2}
\end{figure}
\end{center}
\end{widetext}

\subsection{differential cross section} \label{dcr}

In Fig. \ref{fig-2}, the differential cross sections together with
the partial differential cross sections for several individual
resonances are shown at different c.m. energies from threshold
$W=1.488$ GeV to $W=1.586$ GeV. The experimental
data~\cite{exp1,exp3,exp4,exp6,exp7} are also included for a
comparison.

From the figure, we can see that the calculation results agree
well with the data as shown by the solid curves. The
$S_{11}(1535)$ governs the differential cross sections from the
$\eta N$ threshold to $W=1.586$ GeV, as indicated by the straight
lines in Fig. \ref{fig-2} (1b-12b).

The $S_{11}(1650)$ has significant destructive interferences with
the $S_{11}(1535)$ in the region of $W\leq 1.586$ GeV [see the
dash-dotted curves in Fig. \ref{fig-2} (1b-12b)].

If we switch off the $D_{13}(1520)$, as illustrated by the dashed
curves in Fig. \ref{fig-2} (7a-12a), we find that the shape of the
differential cross sections changes significantly. It shows that
the interference between $D_{13}(1520)$ and $S_{11}(1535)$ are
crucial to produce the correct shape for the differential cross
section around the $\eta N$ threshold. This feature is mentioned
in \cite{1535,aa}, and similar feature also appears in
photoproduction
reactions~\cite{Tiator:1994et,Li:1998ni,qk4,Chiang:2001as,Tiator:1999gr}.

The nucleon pole term contributions are visible in the
differential cross sections [see the dash-dotted curves Fig.
\ref{fig-2} (1a-12a)]. Due to its interference, the differential
cross sections are enhanced in the region of $W\lesssim 1.53$ GeV,
and suppressed in the region of $W\gtrsim 1.54$ GeV by the nucleon
pole.

To see the effects from the $t$-channel, we also show the
differential cross sections without the contributions of it, which
are denoted by the dashed curves in Fig. \ref{fig-2} (1b-12b). In
the region of $W<1.586$ GeV , we find that the contributions from
$t$-channel are very small. Basically, its effects on the
differential cross sections are negligible in this region.

There are nearly no contributions from $D_{13}(1700)$,
$D_{15}(1675)$ and $n=2$ shell resonances for their large
Breit-Wigner masses and /or very small $g_R$ factors. If we switch
off their contributions, the changes of the differential cross
sections are nearly invisible, thus, we do not show them in Fig.
\ref{fig-2}.

Above $W=1.60$ GeV, contributions of the $P$ and $F$-wave
resonances from $n=2$ shell are present, which will be discussed
in Sec. \ref{n2}.

In brief, in the region of $W\lesssim 1.60$ GeV the resonance
$S_{11}(1535)$ governs the process; $D_{13}(1520)$ and
$S_{11}(1650)$ play crucial roles in the reactions; the
contributions of nucleon pole ($s$+$u$-channel) are visible; the
contributions from other resonances and $t$-channel to
differential cross sections are rather small.

\begin{center}
\begin{figure}[bt]
\centering \epsfxsize=9.0 cm \epsfbox{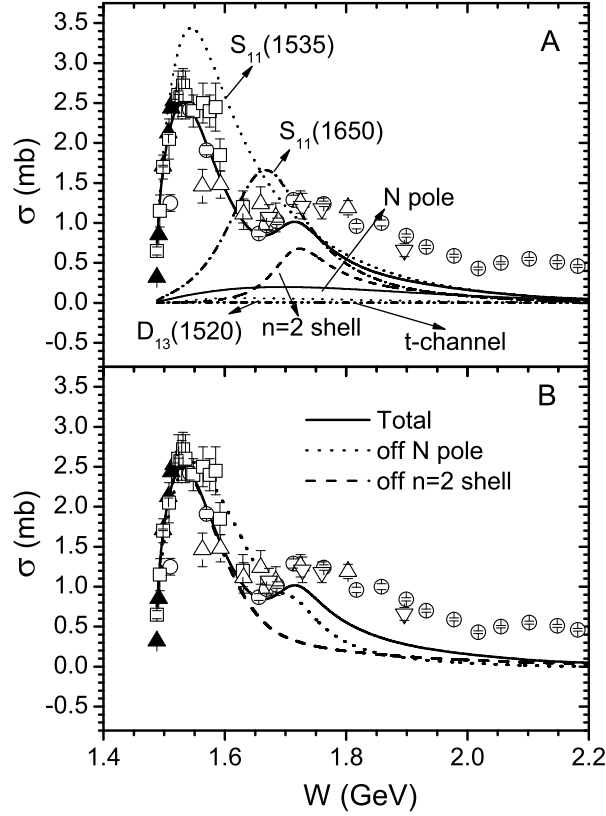} \caption{ The
cross section as a function of $W$. The data are from \cite{exp1}
(open circles), \cite{exp2} (open up-triangles), \cite{exp5} (open
down-triangles), \cite{Clajus:1992dh} (open squares), and the
recent experiment \cite{exp7} (solid triangles). The solid curves
correspond to the full model result. In A, the partial cross
sections for $S_{11}(1535)$, $S_{11}(1650)$, $D_{13}(1520)$, $n=2$
shell and nucleon pole are indicated by different lines and
labelled by corresponding text, respectively. In B, the dotted and
dashed curves are for the results switched off the contributions
from nucleon pole and $n=2$ shell resonances,
respectively.}\label{fig-3}
\end{figure}
\end{center}
\begin{center}
\begin{figure}[ht]
\centering \epsfxsize=9.0 cm \epsfbox{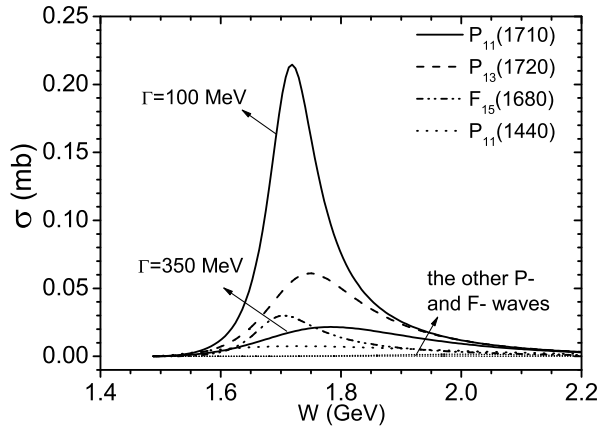} \caption{The
partial cross sections for the resonances in $n=2$ shell are
shown.  All the Breit-Wigner masses and widths for the resonances
are taken from the PDG values. For the $P_{11}(1710)$, the result
with a broader width $\Gamma=350$ MeV is also shown. In the $n=2$
shell, only $P_{11}(1710)$, $F_{15}(1680)$ and $P_{13}(1720)$
contribute to the cross sections obviously. The other resonances,
such as, $P_{13}(1990)$ and $P_{11}(1440)$ nearly have no
contributions to the cross sections. }\label{fig-4}
\end{figure}
\end{center}
\begin{center}
\begin{figure}[ht]
\centering \epsfxsize=7 cm \epsfbox{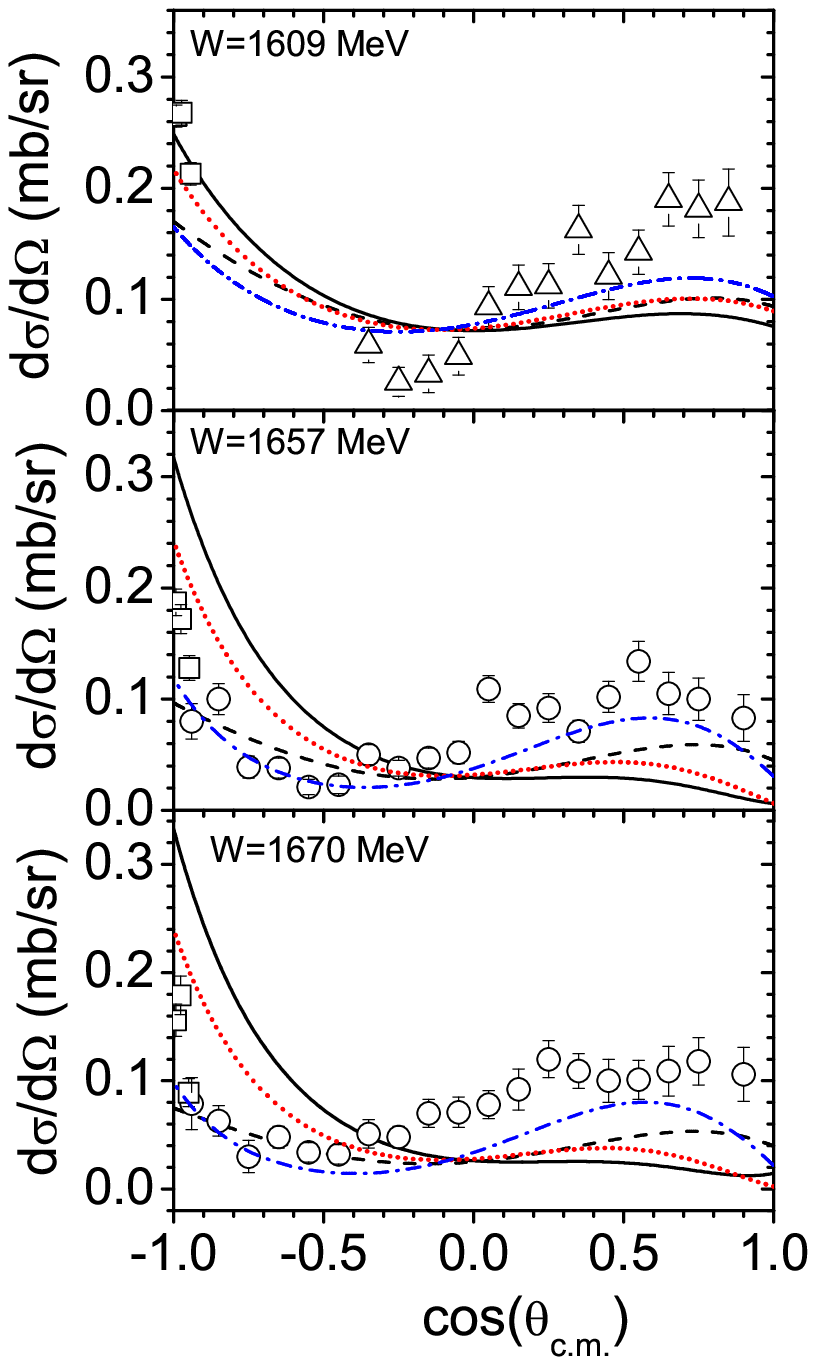} \caption{ (Color
online) The differential cross sections for $W=1.609$, 1.657 and
1.670 GeV, respectively (solid curves). Data are obtained from
\cite{exp1} (open circles). The dashed curves are for the
prediction without the $n=2$ shell resonances. The dash-dotted
curves are for the case when we reverse the sign of the partial
amplitude for $P_{11}(1710)$.}\label{fig-5}
\end{figure}
\end{center}

\subsection{total cross section} \label{crcr}

The total cross section as a function of the c.m. energy $W$ is
plotted in Fig. \ref{fig-3}. To see the contributions of each
resonance, the partial cross sections of a single resonance are
also shown in the same figure. It shows that our theoretical
calculations are in a reasonably good agreement with the
experimental data up to $W \simeq 1.7$ GeV. At higher energies,
although our model gives the correct trend, it underestimates the
total cross section. Interestingly, a ``second peak" around $W\sim
1.7$ GeV appears in the total cross section, which is also
predicted by other models \cite{th5,Penner:2002ma,Shklyar:2004dy}.

Around the threshold, $W<1.6$ GeV (i.e., $p_\pi< 0.9$ GeV), we can
see that the major contributions to the cross sections are from
the $S_{11}(1535)$ and $S_{11}(1650)$. The contributions of the
$S_{11}(1535)$ is about an order of magnitude larger than those
from the  $P$, $D$ and $F$ wave resonances. In this region, it
shows that the exclusive cross section from $S_{11}(1535)$ is even
larger than the data. But the destructive interferences from the
$S_{11}(1650)$ bring down the cross sections.

For $W > 1.6$ GeV, the contributions of $n=2$ resonances appear in
the reaction.  They play important roles around $W=1.7$ GeV.
Without the contributions from $n=2$ shell, the ``second peak"
disappears. To know which resonance in $n=2$ shell contributes to
the ``second peak", we should rely on partial wave analysis. It
will be discussed in Sec.~\ref{n2} later.

There are nearly no contributions from $D_{13}(1700)$,
$D_{15}(1675)$ and $D_{13}(1520)$ in the whole energy region. We
should emphasize that, although there are less contributions of
$D_{13}(1520)$ to the total cross sections, it plays important
roles in the reactions to give a correct shape for the
differential cross sections.

From the exclusive cross section of $t$-channel, we find that the
$t$-channel are negligible to the cross section as shown in Fig.
\ref{fig-3}-A.

Switching off the the contributions from the nucleon pole terms,
we find that the total cross section changes by less than 20\% in
in the region of $W\lesssim 1.6$ GeV, however it decreases
significantly in the region of $W>1.7$ GeV (see the
dash-dot-dotted curve in Fig. \ref{fig-3} B).

A recent analysis of $\pi^- p\rightarrow \eta n$ data suggests the
need of the $P_{11}(1710)$ resonance \cite{shk,Ceci:2006ra}. In
the following subsection, we will discuss those higher resonance
contributions briefly.

\subsection{higher resonances  from $n=2$ shell} \label{n2}

From the analysis in Sections \ref{dcr}, \ref{crcr}, we infer that
when the c.m. energy $W< 1.6$ GeV, the data can be accounted for
with the resonances of $n\leq 1$. To clarify the role played by
the higher resonances, i.e., the $P$ and $F$-wave states in $n=2$
shell, we make an analysis of the differential cross sections in
the energy region $W>1.6$ GeV, where the $P_{11}(1710)$,
$P_{13}(1720)$ and $F_{15}(1680)$ may become important.

Firstly, to see the contributions from individual resonances
[i.e., $P_{11}(1440)$, $P_{11}(1710)$, $P_{13}(1720)$,
$P_{13}(1900)$, $P_{11}(2100)$, $F_{15}(1680)$, $F_{15}(2000)$ and
$F_{17}(1990)$] we plot their partial cross sections as function
of energy in Fig. \ref{fig-4}. It shows that the $P_{11}(1710)$ is
dominant over other states around $W\sim 1.6-1.77$ GeV. Although
the contributions of the $P_{13}(1720)$ and $F_{15}(1680)$ are
visible, they are about $5\sim 10$ times smaller than the
$P_{11}(1710)$. There are nearly no contributions from the
$P_{13}(1900)$, $P_{11}(1440)$ and $F_{17}(1990)$ in $n=2$ in this
energy region. We then conclude that to reproduce the ``second
peak" in Fig. \ref{fig-3} we need the $P_{11}(1710)$, which is
consistent with other studies in the
literatures~\cite{shk,Ceci:2006ra}.

In Fig. \ref{fig-5}, the differential cross sections at $W=1.609$,
1.657 and 1.670 GeV are presented. It shows that without the
$P_{11}(1710)$, $F_{15}(1680)$ and $P_{13}(1720)$, the changes to
the differential cross section are rather significant. We find
that the theoretical predictions overestimate the cross sections
at backward angles, while underestimate the cross sections at
forward angles, compared with the data. Since there are still
large uncertainties with the width of the $P_{11}(1710)$ (i.e.
$\Gamma=50 \sim 450 $ MeV) \cite{PDG,Penner:2002ma}, we thus
adjust it to examine the model predictions. By setting width as
$\Gamma=350$ MeV, we find that the predictions at $W=1.657$ and
1.670 GeV are improved obviously (see the dotted curves in Fig.
\ref{fig-5}). It should be noted that with $\Gamma=350$ MeV for
the $P_{11}(1710)$, its partial cross sections decrease
significantly, and its contributions becomes comparable with those
of $P_{13}(1720)$ and $F_{15}(1680)$ (see Fig. \ref{fig-4}).
Although the predictions are improved by using a broader width for
$P_{11}(1710)$, there still exists a big gap between the
theoretical predictions and the data.

Interestingly, the data seem to favor that the contribution from
the $P_{11}(1710)$ has a reversed sign as shown by the dash-dotted
curves. It also improves the parameter fitting. This could be a
signal for the breakdown of the $SU(6)\otimes O(3)$ symmetry
within the $P$-wave states. A similar example is the radial
excited $P_{11}(1440)$ of $[{\bf 56, \ ^2 8}]$ which is lighter
than the first orbital excited $S_{11}(1535)$ and suggests the
breakdown of the non-relativistic constituent quark model (NRCQM)
~\cite{isgur-karl-model}.

It has also been discussed in the literature that the
$P_{11}(1710)$ could be a candidate for the $1/2^+$ pentaquark
with hidden strangeness~\cite{jaffe-wilczek}. It was shown in
ref.~\cite{zhao-close-2006} that a possible mixture of the $[{\bf
20, \ ^2 8}]$ within the $P$-wave states can break down the naive
quark model symmetry and make their properties very different from
the NRCQM expectations. Our present study certainly does not allow
us to conclude the nature of the $P_{11}(1710)$. But the results
seem to show that the data favor a strong $P$-wave contribution
with a reversed sign in respect of the $P_{11}(1710)$ around
$W\sim 1.7$ GeV, for which the source should be investigated.
Polarization observables in this energy region may be sensitive to
its interference and a partial wave analysis of data should be
pursued.

\section{Summary}\label{sum}

We have extended the chiral quark model approach for meson
photo-production on nucleon to the study of meson-production in
meson-nucleon scatterings. An major advantage of this approach is
that the number of free parameters will be greatly reduced in the
quark model as the leading order calculation. For the reaction
$\pi^{-}p \rightarrow \eta n$ at low energies, we succeed in
accounting for the differential and total cross sections from
threshold to the third resonance region.

In this study, we find that the $S_{11}(1535)$ and $S_{11}(1650)$
dominate the reaction in a wide energy region above the threshold.
Although contributions from  the $D_{13}(1520)$ and nucleon pole
terms are relatively small near threshold, they are crucial to
produce the correct shape of the differential cross sections via
interferences. In particular, the $S_{11}(1650)$ has a destructive
interference with the $S_{11}(1535)$ near threshold, and the
$D_{13}(1520)$ is crucial to produce the angular distributions.
The $t$-channel contributions are negligible in the reactions.
Above the c.m. energy $W\sim 1.6$ GeV, the contributions of higher
resonances from $n=2$ shell also appear. The $P_{11}(1710)$ plays
an important role around the c.m. energy $W=1.7$ GeV, which
contributes to the bump around $W=1.7$ GeV in the total cross
section. It turns out that a sign change for the $P_{11}(1710)$
will better account for the data. This could be a sign for the
breakdown of the NRCQM and state mixings are needed. It may also
be a signal of unconventional configurations inside the
$P_{11}(1710)$ for which both improved experimental measurement
and theoretical phenomenology are required.

\section*{  Acknowledgements }

This work is supported, in part, by the National Natural Science
Foundation of China (Grants 10675131, 10775145), Chinese Academy
of Sciences (KJCX3-SYW-N2), the U.K. EPSRC (Grant No.
GR/S99433/01), the Post-Doctoral Programme Foundation of China,
and K. C. Wong Education Foundation, Hong Kong.


\begin{thebibliography}{99}

\bibitem{Liu:2005pm}
  B.~C.~Liu and B.~S.~Zou,
  Phys.\ Rev.\ Lett.\  {\bf 96}, 042002 (2006)
  [arXiv:nucl-th/0503069].
%
\bibitem{eta-mesic} Q. Haider and L. C. Liu, Phys. Lett. B \textbf{172}, 257 (1986); L. C.
Liu and Q. Haider, Phys. Rev. C \textbf{34}, 1845 (1986). 
\bibitem{Baru:2006hy}
  V.~Baru {\it et al.},
  arXiv:nucl-th/0610011.



\bibitem{exp1} R. M. Brown et al., Nucl. Phys. B \textbf{153},  89
(1979).

\bibitem{exp2} F. Bulos et al., Phys. Rev. \textbf{187}, 1827 (1969).

\bibitem{exp3} W. Deinet, H. Mueller, D. Schmitt, H. M. Staudenmaier, S.
Buniatov, E. Zavattini, Nucl. Phys. B \textbf{11}, 495 (1969).

\bibitem{exp4} J.Feltesse et al., Nucl. Phys. B \textbf{93}, 242 (1975).

\bibitem{exp5} W. B. Richards et al.: Phys. Rev. D \textbf{1},  10 (1970).

\bibitem{exp6} N.~C.~Debenham {\it et al.}, Phys.\ Rev.\  D {\bf 12}, 2545
(1975).



\bibitem{Clajus:1992dh}
  M.~Clajus and B.~M.~K.~Nefkens,
  PiN Newslett.\  {\bf 7}, 76 (1992).

\bibitem{exp7} S. Prakhov et al.: Phys. Rev. C \textbf{72}, 015203
(2005).

\bibitem{Penner:2002ma}
  G.~Penner and U.~Mosel,
  Phys.\ Rev.\  C {\bf 66}, 055211 (2002)
  [arXiv:nucl-th/0207066].



\bibitem{th2}
  R.~A.~Arndt, W.~J.~Briscoe, T.~W.~Morrison, I.~I.~Strakovsky, R.~L.~Workman and A.~B.~Gridnev,
  Phys.\ Rev.\  C {\bf 72}, 045202 (2005)
  [arXiv:nucl-th/0507024].

\bibitem{th3}
  R.~A.~Arndt, W.~J.~Briscoe, I.~I.~Strakovsky, R.~L.~Workman and M.~M.~Pavan,
  Phys.\ Rev.\  C {\bf 69}, 035213 (2004)
  [arXiv:nucl-th/0311089].

\bibitem{th4}
  A.~M.~Gasparyan, J.~Haidenbauer, C.~Hanhart and J.~Speth,
  Phys.\ Rev.\  C {\bf 68}, 045207 (2003)
  [arXiv:nucl-th/0307072].

\bibitem{th5}
  M.~Batinic, I.~Slaus, A.~Svarc and B.~M.~K.~Nefkens,
  Phys.\ Rev.\  C {\bf 51}, 2310 (1995)
  [Erratum-ibid.\  C {\bf 57}, 1004 (1998)]
  [arXiv:nucl-th/9501011].



\bibitem{Shklyar:2004dy}
  V.~Shklyar, G.~Penner and U.~Mosel,
  Eur.\ Phys.\ J.\  A {\bf 21}, 445 (2004)
  [arXiv:nucl-th/0403064].



\bibitem{th6}
  T.~Feuster and U.~Mosel,
  Phys.\ Rev.\  C {\bf 58}, 457 (1998)
  [arXiv:nucl-th/9708051].

\bibitem{Vrana:1999nt}
  T.~P.~Vrana, S.~A.~Dytman and T.~S.~H.~Lee,
  Phys.\ Rept.\  {\bf 328}, 181 (2000)
  [arXiv:nucl-th/9910012].




\bibitem{th1}
  C.~Hanhart,
  Acta Phys.\ Slov.\  {\bf 56}, 193 (2005)
  [arXiv:nucl-th/0511045].


\bibitem{qk1}
  T.~Abdullah and F.~E.~Close,
  Phys.\ Rev.\  D {\bf 5}, 2332 (1972).




\bibitem{qk2}
  F.~E.~Close and Z.~P.~Li,
  Phys.\ Rev.\  D {\bf 42}, 2194 (1990).

\bibitem{qkk} Z. P. Li, Phys.\ Rev.\  D {\bf 48}, 3070 (1993).

\bibitem{qkk0} Z. P. Li, Phys.\ Rev.\  D {\bf 50}, 5639 (1994).

\bibitem{qkk1} Z. P. Li, Phys.\ Rev.\  C {\bf 52}, 1648 (1995).

\bibitem{qkka} Z. P. Li, Phys.\ Rev.\  D {\bf 52}, 4961 (1995).

\bibitem{Li:1997gda}
  Z.~P.~Li, H.~X.~Ye and M.~H.~Lu,
  Phys.\ Rev.\  C {\bf 56}, 1099 (1997)
  [arXiv:nucl-th/9706010].



\bibitem{qkk2} Q. Zhao, Phys.\ Rev.\  C {\bf 64}, 052201(R) (2001).


\bibitem{qk3}
  Q.~Zhao, J.~S.~Al-Khalili, Z.~P.~Li and R.~L.~Workman,
  Phys.\ Rev.\  C {\bf 65}, 065204 (2002)
  [arXiv:nucl-th/0202067].




\bibitem{qk4}
  Q.~Zhao, B.~Saghai and Z.~P.~Li,
  J.\ Phys.\ G {\bf 28}, 1293 (2002)
  [arXiv:nucl-th/0011069].

\bibitem{qk5}
  Q.~Zhao, Z.~P.~Li and C.~Bennhold,
  Phys.\ Rev.\  C {\bf 58}, 2393 (1998)
  [arXiv:nucl-th/9806100].

\bibitem{qk6}
  Q.~Zhao, Z.~P.~Li and C.~Bennhold,
  Phys.\ Lett.\  B {\bf 436}, 42 (1998)
  [arXiv:nucl-th/9803015].

%
\bibitem{isgur-karl-model} N. Isgur and G. Karl, Phys. Lett. {\bf 72B},
109 (1977); N. Isgur and G. Karl, Phys. Rev. D{\bf 18}, 4187
(1978); {\it ibid} D {19}, 2653 (1979); Erratum {\bf 23}, 817
(1981); {\it ibid} D{\bf 20}, 1191 (1979).





\bibitem{glozman}
  L.~Y.~Glozman and D.~O.~Riska,
  Phys.\ Rept.\  {\bf 268}, 263 (1996)
  [arXiv:hep-ph/9505422].

\bibitem{zhangzy-97}
  Z.~Y.~Zhang, Y.~W.~Yu, P.~N.~Shen, L.~R.~Dai, A.~Faessler and U.~Straub,
  Nucl.\ Phys.\  A {\bf 625}, 59 (1997).

\bibitem{huang-05}
  F.~Huang and Z.~Y.~Zhang,
  Phys.\ Rev.\  C {\bf 72}, 024003 (2005)
  [arXiv:nucl-th/0507025].


\bibitem{PDG} W. M. Yao \emph{et al.}, J.\ Phys.\ G {\bf 33}, 1 (2006).
%
\bibitem{aaaa} O. Krehl, C. Hanhart, S. Krewald, and J. Speth, Phys.\ Rev.\  C
{\bf 62}, 025207 (2000).
%
\bibitem{aa} A. M. Gasparyan, J. Haidenbauer, C. Hanhart, and J.
Speth, Phys. Rev C \textbf{68}, 045207 (2003).


%
\bibitem{goldberger-treiman}
  M.~L.~Goldberger and S.~B.~Treiman,
  Phys.\ Rev.\  {\bf 110}, 1178 (1958).


%
\bibitem{Li:1998ni}
  Z.~P.~Li and B.~Saghai,
  Nucl.\ Phys.\  A {\bf 644}, 345 (1998); B. Saghai and Z. Li, Eur. Phys. J. A\textbf{11}, 217 (2001), [nucl-th/0104084];
Proceedings of NSTAR 2002 Workshop on the Physics of Excited
Nucleons, Pittsburgh, Pennsylvania, 9-12 Oct 2002, S.A. Dytman and
E.S. Swanson (Editors), World Scientific (2003) [nucl-th/0305004].


\bibitem{Tiator:1994et}
  L.~Tiator, C.~Bennhold and S.~S.~Kamalov,
  Nucl.\ Phys.\  A {\bf 580}, 455 (1994)
  [arXiv:nucl-th/9404013].

\bibitem{Kirchbach:1996kw}
  M.~Kirchbach and L.~Tiator,
  Nucl.\ Phys.\  A {\bf 604}, 385 (1996)
  [arXiv:nucl-th/9601002].

\bibitem{Zhu:2000eh}
  S.~L.~Zhu,
  Phys.\ Rev.\  C {\bf 61}, 065205 (2000)
  [arXiv:nucl-th/0002018].

\bibitem{Grein:1979nw}
  W.~Grein and P.~Kroll,
  Nucl.\ Phys.\  A {\bf 338}, 332 (1980).






\bibitem{Stoks:1999bz}
  V.~G.~J.~Stoks and T.~A.~Rijken,
  Phys.\ Rev.\  C {\bf 59}, 3009 (1999)
  [arXiv:nucl-th/9901028].

\bibitem{Piekarewicz} J. Piekarewicz, Phys.\ Rev.\  C {\bf 48}, 1555
(1993).



\bibitem{cc2} V. Baru, A. M. Gasparyan, J. Haidenbauer, C. Hanhart, A.
E. Kudryavtsev, and J. Speth, Phys. Rev. C \textbf{67}, 024002
(2003).

\bibitem{cc3} R. Machleidt, K. Holinde, and Ch. Elster Phys. Rep. \textbf{149},
1 (1987).

\bibitem{cc4} R. Machleidt, Adv. Nucl. Phys. \textbf{19}, 189 (1989).

\bibitem{Nasrallah:2005hn}
  N.~F.~Nasrallah,
  Phys.\ Lett.\  B {\bf 645}, 335 (2007)
  [arXiv:hep-ph/0512048].

\bibitem{cc1} M.T. Pe\~{n}a, H. Garcilazo, and D.O. Riska, Nucl. Phys. A\textbf{683}, 322
(2001).



\bibitem{1535} R. A. Arndt, W. J. Briscoe, T. W. Morrison, I. I. Strakovsky, and
R. L. Workman, Phys. Rev C \textbf{72}, 045202 (2005).
%

%
\bibitem{Chiang:2001as}
  W.~T.~Chiang, S.~N.~Yang, L.~Tiator and D.~Drechsel,
  Nucl.\ Phys.\  A {\bf 700}, 429 (2002)
  [arXiv:nucl-th/0110034].
%
\bibitem{Tiator:1999gr}
  L.~Tiator, D.~Drechsel, G.~Knochlein and C.~Bennhold,
  Phys.\ Rev.\  C {\bf 60}, 035210 (1999)
  [arXiv:nucl-th/9902028].
\bibitem{shk} V. Shklyar, H. Lenske, U. Mosel, nucl-th/0611036.



\bibitem{Ceci:2006ra}
  S.~Ceci, A.~Svarc and B.~Zauner,
  Phys.\ Rev.\ Lett.\  {\bf 97}, 062002 (2006)
  [arXiv:hep-ph/0603144].


%
\bibitem{jaffe-wilczek} R. Jaffe and F. Wilczek, Phys. Rev. Lett.
{\bf 91}, 232003 (2003) [arXiv:hep-ph/0307341].
%
\bibitem{zhao-close-2006} Q. Zhao and F.E. Close, Phys. Rev. D
{\bf 74}, 094014 (2006).

\end{thebibliography}
\end{document}